\documentclass{article}
\usepackage[latin9]{inputenc}
\usepackage{textcomp}
\usepackage{amsmath}
\usepackage{graphicx}

\makeatletter

\providecommand{\tabularnewline}{\\}

\usepackage{spconf}

\title{WHISPERED AND LOMBARD NEURAL SPEECH SYNTHESIS}
\name{Qiong Hu, Tobias Bleisch, Petko Petkov, Tuomo Raitio, Erik Marchi, Varun Lakshminarasimhan}
\address{Apple}
\makeatother

\begin{document}
\maketitle 
\begin{abstract}
It is desirable for a text-to-speech system to take into account the
environment where synthetic speech is presented, and provide appropriate
context-dependent output to the user. In this paper, we present and
compare various approaches for generating different speaking styles,
namely, normal, Lombard, and whisper speech, using only limited data.
The following systems are proposed and assessed: 1) Pre-training and
fine-tuning a model for each style. 2) Lombard and whisper speech
conversion through a signal processing based approach. 3) Multi-style
generation using a single model based on a speaker verification model.
Our mean opinion score and AB preference listening tests show that
1) we can generate high quality speech through the pre-training/fine-tuning
approach for all speaking styles. 2) Although our speaker verification
(SV) model is not explicitly trained to discriminate different speaking
styles, and no Lombard and whisper voice is used for pretrain this
system, SV model can be used as style encoder for generating different
style embeddings as input for Tacotron system. We also show that the
resulting synthetic Lombard speech has a significant positive impact
on intelligibility gain.
\end{abstract}

\section{Introduction}

\label{sec:intro}

As the capabilities of digital assistants grow, their use for everyday
tasks is expected to increase significantly. These assistants can
be invoked in a variety of environments. A user may expect a digital
assistant to not only adapt to the level of noise in their current
environment, but also adapt to the current social context by adjusting
the style of speech used. For example, in a quiet library or meeting
room, the user may prefer their interaction with the digital assistant
to be conducted in a more confidential and private way. Whisper is
a widely used style of speech that allows a speaker to limit their
discourse only to nearby listeners, which could be used in this scenario.
Conversely, in a noisy environment, like a cafeteria or driving in
a car, the output speech from devices can be difficult to hear, and
the user may prefer more intelligible speech. Using Lombard speech
could improve sentence intelligibility. Automatic volume adjustment
techniques could be also used in this scenario, but Lombard speech
allows us to increase intelligibility in a human-like fashion, enhancing
the rapport between human and machine, and facilitating interaction.
Therefore, the ability for a digital assistant to generate different
speaking styles with respect to the user's environment would greatly
enhance the user experience in the areas of privacy and intelligibility,
in addition to improving the naturalness of interactions.

Neural TTS system based on end-to-end models \cite{wang2017tacotron,adiga2018use,oord2016wavenet,oord2017parallel,shen2018natural,ping2018deep,arik2017deep}
have shown the ability to generate high-quality synthetic speech.
However, most of the systems target a single ``normal'' speaking
style, or focus on prosody modeling \cite{skerry2018towards,wang2018style}.
The goal of this paper is to develop a robust text-to-speech system
capable of synthesizing high-quality whisper and Lombard speech in
addition to normal style.

Most existing works on whisper and Lombard speech synthesis analyze
and generate these speaking styles separately. Because whisper speech
lacks the vibration element of voiced speech, the fundamental frequency
(f0) and spectral tilt can be modified in a source-filter model to
convert normal speech into whisper speech \cite{raitiowhisper2016,cotescu2019voice}.
Other methods \cite{toda2007voice,cotescu2019voice} learn the mapping
between normal and whisper acoustic features through voice conversion
(VC) techniques. However, parallel corpora containing both normal
and whisper recordings are usually needed for training a VC model.

Data-driven methods for Lombard speech generation have gained popularity
in recent years \cite{raitio2011lombard,suni2013,cooke2013intelligibility,raitio2014lombard,bollepalli2019b,bollepalli2019lombard}.
These methods typically require a corpus of Lombard speech recordings.
Although neural TTS systems based on end-to-end models \cite{wang2017tacotron,adiga2018use,oord2016wavenet,oord2017parallel,shen2018natural}
can generate high-quality speech, very few studies address Lombard
and whisper speech. Treating these modalities as speaking styles that
can be generated from a single model is another dimension that deserves
further attention.

To investigate the feasibility of synthesizing whisper and Lombard
speech with neural TTS, while keeping in mind the difficulties involved
in data collection for whisper and Lombard speech, we created neural
TTS systems based on the Tacotron 2 \cite{wang2017tacotron} architecture
capable of synthesizing different speaking styles using a limited
amount of data. The data used was collected internally, resulting
in 120 minutes of speech data across three different styles from a
single speaker: normal, whisper, and Lombard. We employ speech adaptation
methods to fine-tune a target style voice on a model trained with
multi-speaker data from an open source multi-speaker dataset. In addition,
a multi-style selection model based on speaker embeddings is also
proposed. For reference, we also leverage signal processing techniques
to convert a generated normal speaking style to both Lombard and whisper
styles as baselines. Results show that we can generate new style-specific
speech through pre-training and fine-tuning while achieving quality
close to that of a natural speech. The neural-adapted whisper speech
achieved higher quality compared to the signal processing based technique.
We further show that we can generate different speaking styles through
a single multi-style model with high quality, even with a database
with uneven amounts of style-specific data.

In the context of Lombard speech, there have been a number of studies
on improving speech intelligibility. Spectral shaping and dynamic
range compression, induced statically or adaptively (by optimizing
an objective measure of intelligibility), have proven highly effective
under various noise conditions \cite{zorila2012speech,Petkov2015}.
We investigate the intelligiblity-enhancing properties of neural-adapted
Lombard speech by comparing it to a post-processing strategy based
on an intelligibility-enhancing algorithm.

This paper is organized as follows. We first introduce the data collection
and preparation in Section 2. Several multi-style model architectures
are explained in Section 3. Experiments and results are shown in Section
4. The discussion and conclusions are presented in Section 5.

\section{Multi-style speech corpus}

\label{sec:format}

\begin{figure}[t]
\centering \includegraphics[scale=0.35]{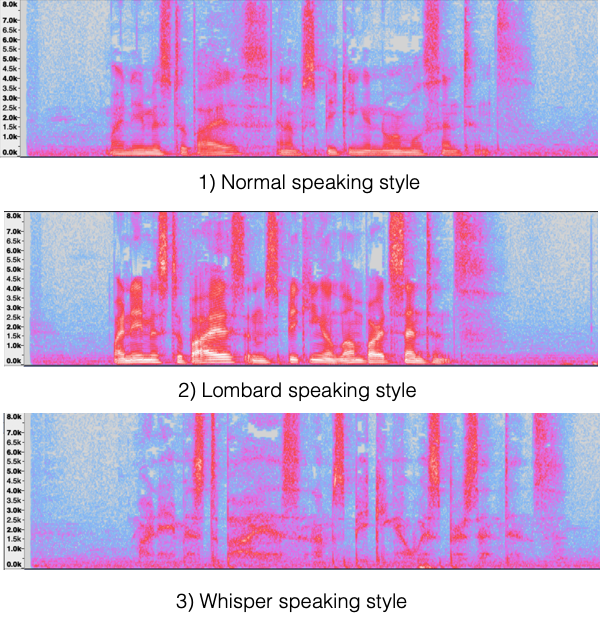} \vspace{0mm}
 \caption{Spectrum of normal (top), Lombard (middle) and whisper speech (bottom)
for the same sentence.}
\label{fig:spectrum} \vspace{0mm}
\end{figure}

The multi-style speech corpus used in this work was created in a professional
recording studio over the course of three days. Normal, Lombard, and
whisper style speech was recorded by an American, non-professional,
male talent. Each style was collected through a 4-hour session (one
style per day), yielding around 40 minutes of usable audio per session.
The talent read out the same 600 sentences per style, most of which
were selected from VCTK \cite{veaux2016superseded} and voice assistant
scripts. The complete dataset consisted of 2 hours (3 $\times$ 40
min) of audio sampled at a frequency of 24~kHz. In this preliminary
investigation, the text and lexicon were not cleaned. The normal style
speech was recorded in clean conditions, whereas the Lombard speech
was recorded while pre-recorded cafeteria noise was played over the
talent's headphones at 65~dB. The whisper style recording is unvoiced
as the talent made an effort to keep the vocal cords not to vibrate.
The visualization of the spectrum of each speaking style for the same
sentence is shown in Fig.~\ref{fig:spectrum}.

Meanwhile, 100 speakers from the VCTK corpus were also used for training
a multi-speaker system. The corpus consists of around 40 hours of
audio, with each speaker reading the same script for about 20 minutes.
To supplement our speech data for training the universal vocoder for
each style, we made use of additional corpora. For whisper speech,
we used the CSTR whisper TIMIT Plus corpus \cite{yang2012noise},
which contains 421 sentences selected from newspaper text. For Lombard
speech, we used the Hurricane Natural Speech Corpus published by the
LISTA Consortium \cite{cooke2013intelligibility}, which contains
720 sentences selected from the Harvard collection. For normal speech,
we used our internal dataset of an American English male speaker,
from which we selected about 750 utterances for this work.

\section{Architecture for multi-style models}

\label{sec:pagestyle}

Our baseline neural TTS system is comprised of two networks. The frontend
is an attention-based sequence-to-sequence network, similar to Tacotron~2
\cite{wang2017tacotron}, which takes phoneme sequence together with
punctuation and word boundaries as input, and generates a Mel-spectrogram
as output. The backend is a single-layer recurrent neural network,
similar to WaveRNN \cite{kalchbrenner2018efficient}, which is applied
as a vocoder for generating the waveform from the Mel-spectrogram.
In the following sections, we propose different system architectures
for generating voices with multiple speaking styles.

\subsection{Pre-training/fine-tuning adaptation model (PF model)}

\label{sec:pf}

Although our neural-net-based end-to-end (E2E) TTS baseline can generate
high quality speech with the appropriate prosody, building an E2E
TTS system usually requires a large amount of data for each style.
In \cite{hu2019neural}, we showed that transfer learning, i.e. pre-training
on a multi-speaker dataset as the initialization, and then fine-tuning
the model on a target speaker's voice, can significantly decrease
the amount of data needed while maintaining the quality and speaker
similarity of the generated speech.

Similarly, in low-resource environments, the pre-training/fine-tuning
model \cite{bengio2009curriculum,vilalta2002perspective} is utilized
to adapt to each style. In the pre-training process, a multi-speaker
Tacotron using only linguistic features as input is trained to learn
a general text-to-speech task. During the fine-tuning stage, each
style is considered as an unseen speaker to adapt the Tacotron weights
to the target style.

\subsection{Post-processing approach}

Another popular method to generate different speaking styles is to
apply handcrafted signal processing (SP) methods (as a post-filter)
to convert normal phonated speech to Lombard or whisper speech. This
approach offers the advantage of synthesizing any style without the
need for data collection for the target style.

In order to have a reference on how well the whisper and Lombard adaptation
works, we used a SP-based conversion technique as a baseline. For
whisper voice, the method is based on frame-wise source-filter processing
of the input speech signal using linear prediction \cite{raitiowhisper2016}.
The speech signal is inverse filtered, and the residual signal is
replaced with an energy-scaled noise signal in order to imitate whisper.
Also, the overall spectrum, as well as formant frequencies and bandwidths,
are modified to resemble whispered speech.

\begin{figure}[t]
\centering\includegraphics[scale=0.15]{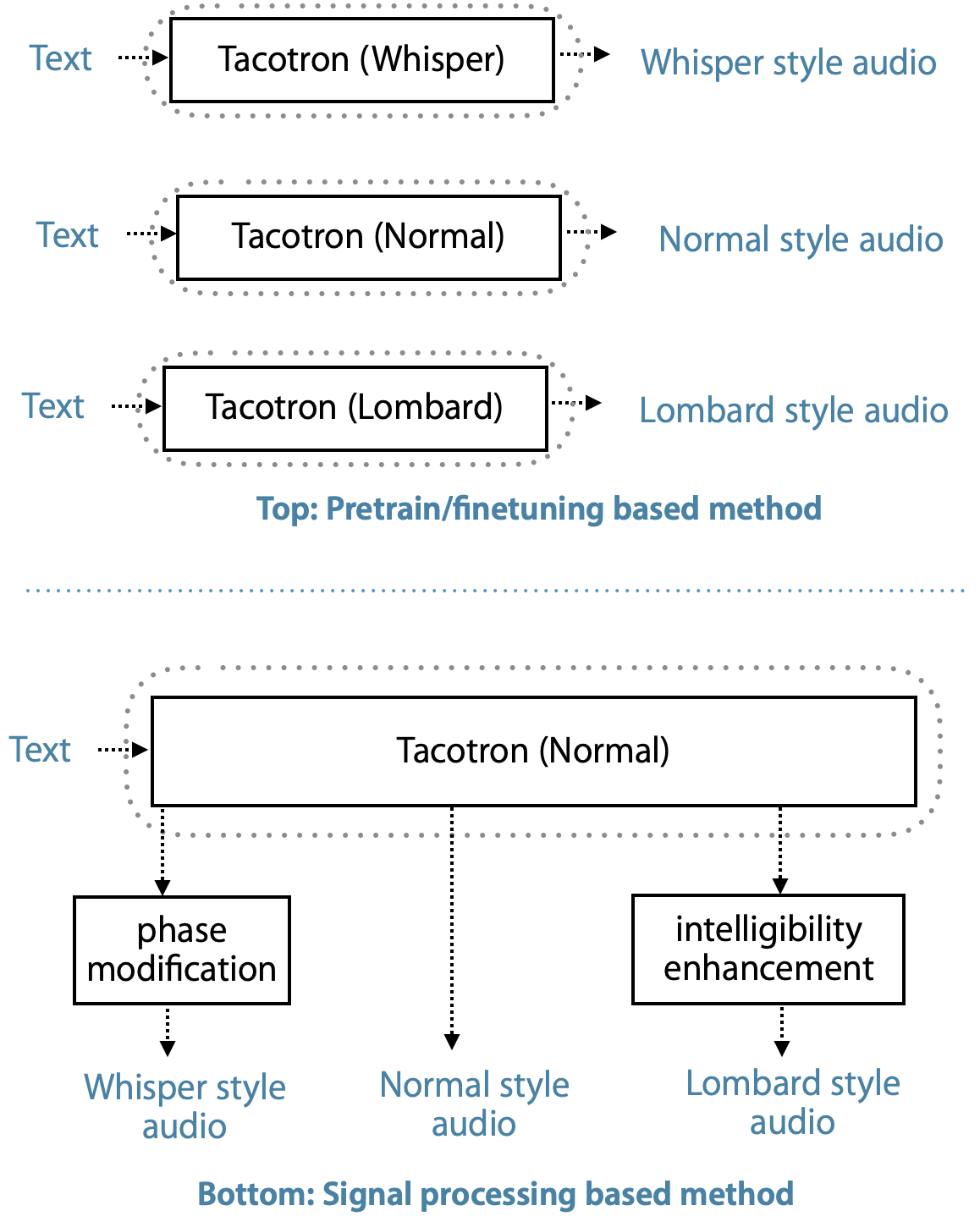}\vspace{0mm}
 \caption{Systems for generating different speaking styles. Top: pre-training/fine-tuning
adaptation model. Bottom: post-processing model.}
\label{fig:post-proces} \vspace{0mm}
\end{figure}

To generate intelligibility-enahnced speech through post-processing,
we used an enhanced internal implementation of \cite{Petkov2015}.
The method was selected based on its performance characteristics and
ability to adapt to the speech signal statistics as well as the noise.
A block diagram depicting the differences between PF and post-processing
approaches is shown in Fig.~\ref{fig:post-proces}.

\subsection{Multi-style model using speaker embedding}

Although the PF model can generate the target voice from a limited
amount of data, the current architecture requires training a separate
model for each style. \cite{jia2018transfer,chen2018sample,Wei2018nips}
have proposed extending the Tacotron to a multi-speaker model by conditioning
the synthesis on a speaker representation in addition to the phoneme
sequence. \cite{zhang2019learning} further extends the model to a
multi-speaker, multi-language model by using language embedding as
an additional input. For easy control, we propose a multi-style, multi-speaker
Tacotron model for enabling the generation of multiple styles from
a single model.

It is important to note that, although our own voice talent recorded
speech in all the 3 styles, other speakers in the multi-speaker corpus
(VCTK) use only normal speaking style. The percentage of Lombard and
whisper speech data is limited, and the training data per style is
unbalanced. Therefore, instead of adding an additional encoder for
conditioning the speaking style as in \cite{wang2018style,wang2017tacotron,cooper2020zero},
we treat normal, Lombard and whisper speech from the target talent
as 3 different speakers, combining them together with the multi-speaker
dataset for training.

In order to extract speaker information we used a pre-trained speaker
encoder based on the speaker verification (SV) system in \cite{siriteam2018PHS,marchi2018generalised}.
The speaker encoder used in this work is trained in a similar way
as described in \cite{marchi2018generalised} but with a slightly
different model architecture. 

A sequence of MFCC frames (20 MFCCs per frame, 25 ms data window,
100 frames per second) is passed as in- put through a 2-layer LSTM
with 512 units each, followed by a self-attention mechanism, a speaker
embedding layer with 128 units and a softmax layer with 18k units.
The self- attention mechanism operates on the outputs from the last
LSTM layer and computes three L\texttimes D projections namely query
(Q), key (K) and value (V) where L denotes the num- ber of frames
and D is the dimension of the LSTM output. We compute a dot product
between Q and K, apply a row- wise softmax and summarize the output
using a row-wise max operation. This way we obtain an L\texttimes 1
dimensional context vector that is then multiplied by the L\texttimes D
LSTM output and subsequently passed to the speaker embedding layer.

\begin{figure}[t]
\centering \includegraphics[scale=0.28]{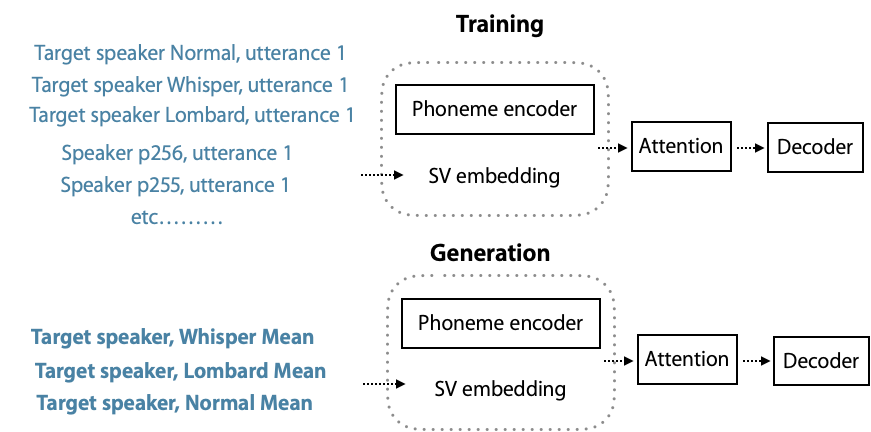} \vspace{0mm}
 \caption{Architecture of multi-speaker, multi-style model.}
\label{fig:SV} \vspace{0mm}
\end{figure}

During training, utterance-level speaker embedding features together
with the corresponding phoneme sequence and Mel-spectrogram are used
to train the multi-speaker Tacotron. In the generation phase, style-level
embedding features (mean value for each style from the target speaker)
are precomputed and applied to control whether the generated voice
is normal, Lombard, or whisper. The architecture of the multi-style,
multi-speaker model is shown in Fig.~\ref{fig:SV}.

\begin{figure}[t]
\centering \includegraphics[scale=0.6]{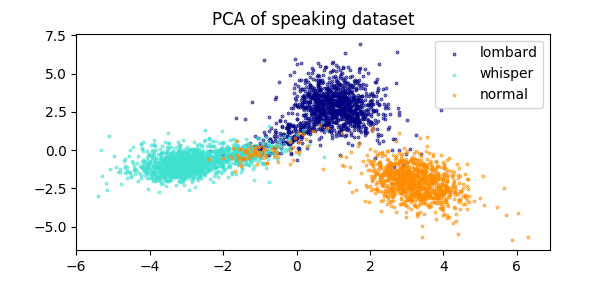} \vspace{0mm}
 \caption{PCA of the SV embedding for each utterance of Lombard (blue), whisper
(cyan), and normal (orange) speech from the target speaker. The same
utterances are read for each style.}
\label{fig:PCA}
\end{figure}

We visualize style embeddings from the speaker encoder with principal
component analysis (PCA) \cite{wold1987principal} for each utterance.
As the SV model is trained to distinguish between speakers, embeddings
from the same speaker which have high cosine similarity should be
distributed closely as a cluster in the speaker space. However, we
can see from Fig.~\ref{fig:PCA} that even for a single speaker,
the Lombard, normal, and whisper embeddings have formed a distinct
cluster for each style. Therefore, we can draw the following conclusions:
1) speaker embeddings represent rich information about speaking styles
as well as speaker identity, 2) although there is no Lombard and whisper
style dataset existing for training the SV model, and the SV model
is not explicitly trained to discriminate different speaking styles,
we can use the speaker embedding features from Lombard, normal and
whisper speech as speaking style representations to control synthetic
speech. No additional speaking style encoder is added to condition
the Tacotron. The cluster of different styles can help in future studies
on copying unseen speaking styles between speakers.

\section{Experiments}

\label{sec:typestyle}

The input for the baseline Tacotron model is a sequence of phonemes,
punctuations, and word boundaries, which is converted to 512-dimensional
embeddings through a text encoder. The output is a sequence of Mel-spectrograms,
computed from 25~ms frames with 10~ms shift. A stepwise monotonic
attention \cite{he2019robust} is adopted for better alignment between
the input and output features. For evaluation of the naturalness,
we use a mean opinion score (MOS) on a scale from 1 to 5. AB listening
tests are conducted to assess which system sounds more natural or
more preferred in terms of speech quality or intelligibility (A: first
sample, B: second sample, C: no preference). 30 test utterances are
generated from each system, and each utterance is evaluated by native
listeners using headphones (30 for MOS, at least 15 for AB).

A preliminary test was first conducted to determine whether combining
datasets from distinct speakers performing the same style can further
improve the quality of the WaveRNN for that style. The training data
and preference scores are shown in Table~\ref{tab:waveRNN}. The
results show that there is no strong preference for audio generated
using the universal vocoder (trained on the combined dataset). We
assume that this is due to the limited amount of data used and lower
quality speech in the public corpus compared to the studio dataset.
Therefore, for all systems, the speaker and style-dependent WaveRNN
is used\footnote{Natural and synthetic speech samples can be found at: https://qsvoice.github.io/samples.html}.

\begin{table}[t]
\caption{Results of the AB test comparing WaveRNN models using different data.
A: Using 40 min of target speech for each style. B: Combining target
speech with a dataset of the same style (CSTR Lombard, internal male
dataset, CSTR whisper).}
\label{tab:waveRNN} \centering \vspace{5mm}
\begin{tabular}{|c|c|c|c|}
\hline 
Style & A & B & No preference\tabularnewline
\hline 
Lombard & 18 \% & 18 \% & 64 \%\tabularnewline
Normal & 12 \% & 15 \% & 73 \%\tabularnewline
Whisper & 36 \% & 14 \% & 50 \%\tabularnewline
\hline 
\end{tabular}\vspace{0mm}
\end{table}

\vspace{0mm}

\subsection{Evaluation}

First, we evaluate how well the pre-training and fine-tuning models
perform with a limited amount of target speech data. An average Tacotron
model is pre-trained using the VCTK dataset. Then the initialized
parameters are fine-tuned by the target speaker's normal, Lombard,
and whisper speech, resulting in three models: PF\_Lombard, PF\_whisper
and PF\_normal, respectively. Post\_whisper is the system applying
voice source modification via digital signal processing methods based
on the PF\_normal output as the reference. The MOS quality results
are shown in Table~\ref{tab:MOS}.

\begin{table}[t]
\caption{MOS quality results with natural speech, synthetic speech from pre-training
and fine-tuning model, and whisper speech from post-processing approach.}
\vspace{5mm}
 \label{tab:MOS} \centering %
\begin{tabular}{|c|c|c|}
\hline 
System & Style & MOS\tabularnewline
\hline 
\hline 
Natural speech & Lombard & 4.47\tabularnewline
PF\_Lombard & Lombard & 4.08\tabularnewline
\hline 
\hline 
Natural speech & Normal & 4.60\tabularnewline
PF\_normal & Normal & 4.32\tabularnewline
\hline 
\hline 
Natural speech & Whisper & 4.42\tabularnewline
PF\_whisper & Whisper & 4.26\tabularnewline
Post\_whisper & Whisper & 2.95\tabularnewline
\hline 
\end{tabular}
\end{table}

We observe that natural speech collected from our studio can achieve
very high MOS (around 4.5). With only around 40 minutes of audio,
our neural TTS adaptation models are also able to generate high quality
speech for all three styles (PF\_Lombard, PF\_whisper and PF\_normal)
with MOS scores above 4. The signal processing based system Post\_whisper,
however, resulted in a much lower MOS (below 3). The results imply
that: 1) Our pre-training/fine-tuning models are robust enough to
generate speech with different styles, and with limited amount of
data for each style, we can synthesize speech with quality close to
that of natural speech. 2) Applying the signal processing method enables
us to generate whisper speech without target recordings. However,
when considering overall speech quality, the neural adaptation method
performs better.

Next, we assessed whether we can use the multi-speaker, multi-style
model to synthesize normal, whisper, and Lombard speech from a single
model while maintaining high quality. A multi-speaker, multi-style
Tacotron is trained such that the speaker encoder is pre-trained by
an attention-based speaker verification model to compute the speaker
embedding. The normal, whisper, and Lombard styles are treated as
three additional speakers trained together with the VCTK dataset.
AB tests are conducted to compare their quality to the PF model described
in Sec.~\ref{sec:pf}. From results in Table~\ref{tab:ABX_SV},
we can see that, except for the normal speaking style, the SV-based
system is generally preferred over the PF model. For whisper and Lombard
styles, the SV-based multi-style model can outperform the speaker-dependent
models. We assume that this might be due to the limited amount of
whisper and Lombard speech data from the target speaker and the fact
that the PF model is easily over-tuned for those styles. Nevertheless,
it indicates that we can generate speech with different styles from
a single model with comparable quality to the individual PF models.
Our next step is to investigate whether we can transfer speaking styles
across speakers by SV embedding modification.

\begin{table}[t]
\caption{AB test results comparing the pre-training/fine-tuning (PF) model
and SV based multi-style model.}
\vspace{5mm}
 \label{tab:ABX_SV} \centering %
\begin{tabular}{|c|c|c|c|}
\hline 
Style & PF model & Multi-style model & No pref.\tabularnewline
\hline 
Lombard & 40 \% & 49 \% & 11 \%\tabularnewline
Normal & 52 \% & 39 \% & 9 \%\tabularnewline
Whisper & 36 \% & 46 \% & 17 \%\tabularnewline
\hline 
\end{tabular}\vspace{0mm}
\end{table}

A comparison between two of the speaking styles (Normal and Lombard)
and their intelligibility-enhanced versions was performed to evaluate
how easy to understand these are in the presence of noise. Intelligibility
enhancement was achieved by applying a signal-processing technique
to the TTS output \cite{Petkov2015}. Two subjective experiments were
conducted for a single noise condition: cafeteria noise at $-4$ dB
SNR.

The first experiment was an AB test aimed at obtaining an estimate
of listening effort. Subjects were instructed to give preference to
the version in which speech was more present. The results (averages
over 8 perticipants), shown in Table~\ref{tab:MOS-1-1}, suggest
an advantage for the neural-based PF model for Lombard speech over
the normal PF model with post-processing (PP) for intelligbity enhancement.
We also compared whisper PF and Lombard PF speech to their respective
PP-enhanced versions. In both cases, preference scores favour the
use of the intelligibility-enhancement technique.

\begin{figure}[h]
\centering \includegraphics[scale=0.21]{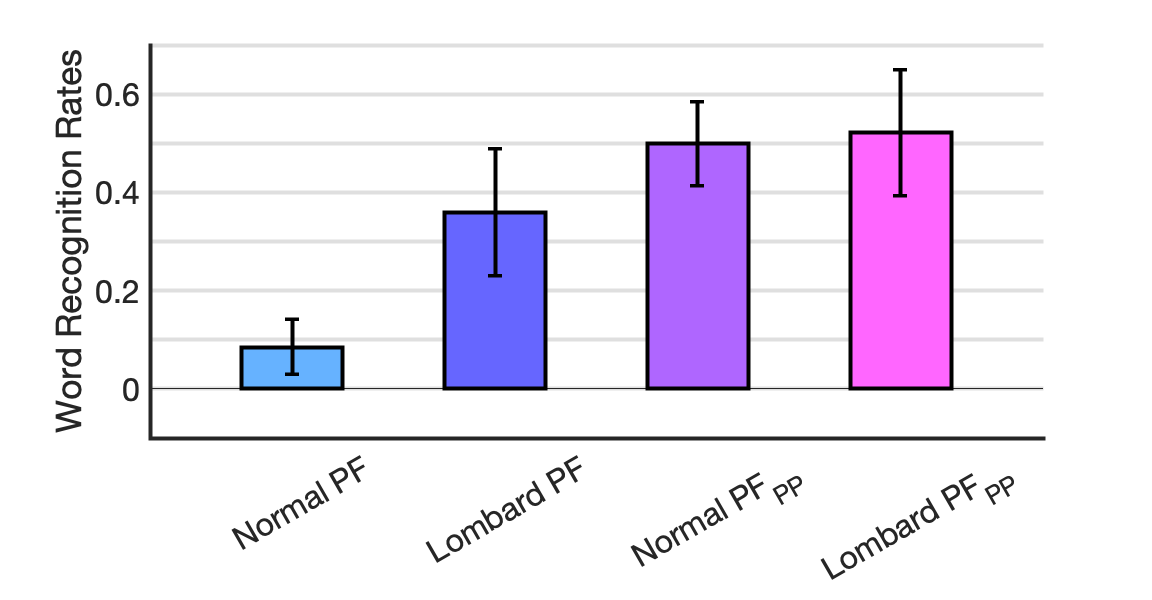} \vspace{0mm}
 \caption{Word recognition rates (and 95\% confidence intervals) from listening
test at $-4$ dB SNR cafeteria noise.}
\label{fig:intel_bars} 
\end{figure}

The second experiment was an intelligibility test from which we compute
word recognition rates (WRR). Speech utterances are mixed with noise
and presented to listeners through headphones. Using an HTML interface,
we ran the test remotely following the protocol from \cite{cooke2013intelligibility}.
We used a subset of the Harvard sentences synthesized with the proposed
neural TTS models. Each modality was represented by 20 utterances
and the system-to-utterance assignements were varied between subjects
(without breaking the sets). Sentences from different systems were
presented with a random order. At the time of reporting, we had received
the results from seven en-US native speakers. The results in Fig.
\ref{fig:intel_bars} show that Lombard speech improves intelligibility
significantly over Normal PF. In turn, post-processed Normal PF (Normal
PF$\mathrm{}_{\mathrm{PP}}$) and post-processed Lombard PF (Lombard
PF$\mathrm{}_{\mathrm{PP}}$) improve over Normal PF.

\begin{table}[t]
\caption{AB test results between pre-training/fine-tuning (PF) model and intelligibility-enhanced
versions (PP) in noise.}
\vspace{5mm}
 \label{tab:MOS-1-1} \centering %
\begin{tabular}{|c|c|c|}
\hline 
Lombard PF & Normal PF$\mathrm{}_{\mathrm{PP}}$ & No preference\tabularnewline
\hline 
47 \% & 35 \% & 18 \%\tabularnewline
\hline 
\hline 
Whisper PF & Whisper PF$\mathrm{}_{\mathrm{PP}}$ & No preference\tabularnewline
\hline 
17 \% & 52 \% & 31 \%\tabularnewline
\hline 
\hline 
Lombard PF & Lombard PF$\mathrm{}_{\mathrm{PP}}$ & No preference\tabularnewline
\hline 
13 \% & 65 \% & 22 \%\tabularnewline
\hline 
\end{tabular}
\end{table}


\vspace{0mm}

\section{Discussion and conclusions}

\label{sec:majhead}

It is desirable for a TTS system to take into account the environment
where synthetic speech is presented. For example, the change of speaking
style to Lombard speech in a noisy environment can make the synthesized
speech more intelligible, and the change to whisper in a quiet environment
can keep the user's conversation with the voice assistant private.
In this paper, we investigated the generation whisper and Lombard
speech using only limited data. We proposed and compared several approaches:
1) Pre-training and fine-tuning a separate model for each style. 2)
Whisper and Lombard speech conversion through a signal processing
based approach. 3) Generation of multiple styles using a single multi-style
model by pre-training the model with a speaker verification model
output. We showed that the pre-training and fine-tuning approach allows
us to generate all the styles using a single model and only 40 minutes
of speech from each style. Our MOS, AB, and WRR listening tests show
that the output speech quality is close to natural speech, and that
the generated Lombard speech has a measurable and significant intelligibility
gain. Our future work includes style transfer between speakers, and
exploring prosody modeling techniques for controlling speaking styles.

\vfill{}

\clearpage{}

\newpage{}

\bibliographystyle{IEEEbib}
\bibliography{Template}

\end{document}